\documentclass[twocolumn,showpacs,preprintnumbers,amsmath,amssymb]{revtex4}

\usepackage[dvips]{graphicx}
\usepackage{dcolumn}
\usepackage{bm}

\begin{document}


\title{Geometrically Frustrated Crystals: Elastic Theory and Dislocations}

\author{Masahiko Hayashi}
 \email{hayashi@cmt.is.tohoku.ac.jp}
\author{Hiromichi Ebisawa}%
\affiliation{%
Graduate School of 
Information Sciences, Tohoku University, 
Aramaki Aoba-ku, Sendai 980-8579, Japan and \\
JST-CREST, 4-1-8 Honcho, Kawaguchi, Saitama 332-0012, Japan
}%
\author{Kazuhiro Kuboki}
\affiliation{Department of Physics, Kobe University, Kobe 657-8501, Japan}

\date{\today}

\begin{abstract}
Elastic theory of ring-(or cylinder-)shaped crystals is 
constructed and the generation of edge 
dislocations due to geometrical frustration caused by the bending 
is studied. 
The analogy to superconducting (or superfluid) vortex state is 
pointed out and the phase diagram of the ring-crystal, 
which depends on radius and thickness, is discussed. 
\end{abstract}

\maketitle
What is the lowest energy state of matter? 
The answer to this fundamental question has been 
believed to be the \lq\lq crystalline state\rq\rq\, 
as far as the system has an infinite extension 
and quantum fluctuation is negligible. 
Most elements or compounds freeze into 
crystals when cooled down slowly. 
Even the electrons form the so-called Wigner crystal if 
the density is low enough. 
The basic character of the crystalline state is the breaking of 
translational symmetry; the system is divided into small fundamental cells. 
Although some exceptions, such as Penrose lattice or 
quasicrystals, have been known, 
this profound form of existence occupies 
the most intriguing expositions in the \lq\lq 
mineralogical department of natural history museum\rq\rq
\cite{Ashcroft-Mermin}. 

In recent decades, with the help of developed 
fine processing technology, especially 
symbolized by the term \lq\lq nanotechnology\rq\rq, 
a new field seems to be added to the 
studies of crystallography. 
The topic of this field is the crystallization in nano-scales. 
A group of materials which is most attracting attention from both 
basic and applied scientists  
is evidently Fullerenes and nanotubes 
\cite{iijima,curl-smally,kroto_review,
iijima_nanotube,saito,hayashi-nanotube}, 
where spherical or cylindrical form of crystals and 
their subspecies are found to be stable structures 
of carbon in nano-scales. 
Tanda et al. have shown that quasi-one-dimensional 
crystals, such as TaS$_3$ or NbSe$_3$, 
which are usually grown as needle-shaped crystals, 
can be made into rings or cylinders by controlling the 
growing conditions\cite{Tanda1,tanda2,Okajima,Tanda,Tsuneta}. 
They have even succeeded in producing single-crystal 
M\"obius strips. 
These crystals with topologically 
nontrivial shapes are named \lq\lq topological matters\rq\rq. 
Some other cylindrical crystals were also found recently, such as 
MoS$_2$\cite{MoS2}. 
The expanding variety of the topological matters 
seems to be invoking questions about 
the nature of crystallization in nano-scales. 

The crystal structure largely affects the physical properties 
of the system and, therefore, 
these crystals provide excellent experimental environments 
in investigating topological physics. 
For example, TaS$_3$ and NbSe$_3$ 
show superconductivity and 
charge density wave order under appropriate conditions
\cite{Gruner}, 
and, then, topology-induced novel states are expected. 
Several theoretical works have been elaborated 
in this direction
\cite{Hayashi-Ebisawa,Hayashi-Ebisawa-Kuboki,
Yakubo-Avishai-Cohen,Wakabayashi-Harigaya}, 
and experimental investigations are under way. 

\begin{figure}[htb]
\includegraphics[width=7cm,clip]{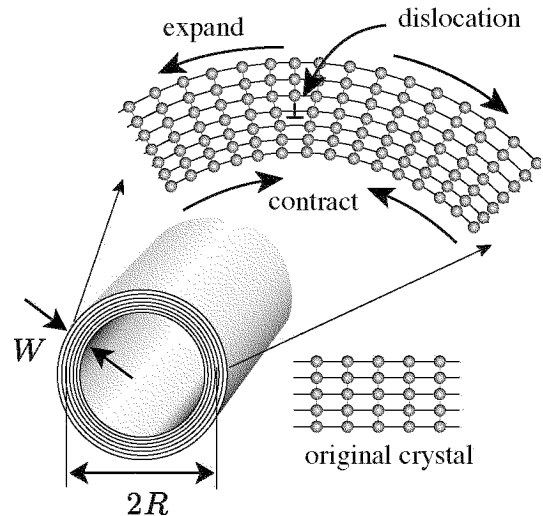}
    \caption[test]{
	The structure of the ring-(or cylinder-)shaped 
	crystal made of a orthorhombic lattice. 
	Because of the expansion and contraction caused by 
	the bending, edge dislocations are likely to be generated. 
	Here $W$ is the thickness of the ring and $R$ is 
	the radius. 
	$R$ is defined at the center of the thickness. 
         }
    \label{dislo}
\end{figure}

In order to understand the physical properties of the 
\lq\lq topological matters\rq\rq, 
the knowledge about crystal structure 
is indispensable. 
In this paper we study the structures of 
ring-(or cylinder-)shaped crystals, 
which we call simply the \lq\lq ring crystals\rq\rq\, 
hereafter. 
Special attention is paid to the formation of 
crystal defects due to the elastic stress 
which topological matters undergo 
because of their nontrivial forms. 
In case of the ring crystals the edge dislocations 
are likely to be generated because of the bending, 
as one can see from Fig. \ref{dislo}. 
In this paper we derive, within the linear elastic theory, 
the condition for the formation of edge dislocations 
in a ring crystal whose radius and thickness are $R$ and $W$, 
respectively. 
Analogy between the superconducting (or superfluid) 
vortices and the dislocations in a ring crystal is invoked. 
Our results may be useful in understanding physical phenomena 
in ring crystals. 

\vspace{2mm}
\noindent
\underline{\it Elastic theory of a ring crystal}

\begin{figure}[htb]
\includegraphics[width=6cm,clip]{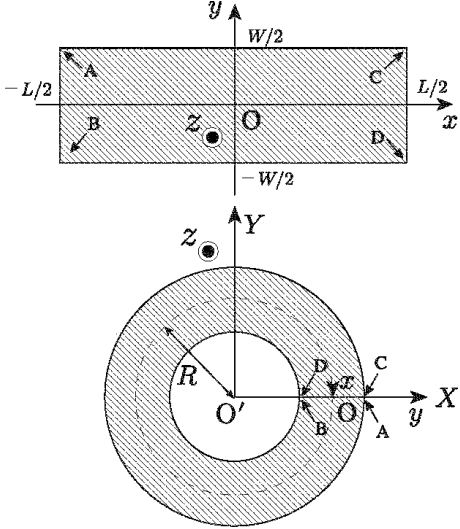}
    \caption[test1]{
	Making a rectangular strip of a crystal into a ring. 
	The strip (a) is made into a ring (b) 
	by introducing the elastic deformation. 
	The system is assumed to be uniform in $z$-axis, 
	which is perpendicular to the page. 
         }
    \label{mapping}
\end{figure}

\vspace{1mm}
To construct the elastic theory of a ring crystal 
we start by making a rectangular strip of a 
crystal into a ring as depicted in Fig. \ref{mapping}, 
which introduces elastic deformation. 
We assume that the system is 
uniform in $z$-direction (along the axis of the 
cylinder) and suppress the $z$-coordinate hereafter. 
The free energy is considered to be that for unit length in 
$z$-direction. 
The crystal is anisotropic 
and it is assumed that the elastic coupling 
in the direction along A-C and B-D is 
much stronger than that along A-B and C-D
(see Fig. \ref{mapping}). 
This situation is realized in most ring crystals 
grown until now.

The elastic free energy of the ring crystal 
is constructed by examining how the linear element 
of the original crystal $d \vec{r} = (dx, dy)$ is 
modified by the bending in the ring crystal. 
Let us suppose that 
the position in the original crystal $\vec{r}=(x,y)$ 
is moved to $\vec{R}=(X,Y)$ in the ring crystal. 
Here orthogonal coordinate system is used. 
We obtain the following relation, 
\begin{align}
(X,Y)=\left((r+v) \cos\frac{x+u}{R}, 
(r+v) \sin\frac{x+u}{R} \right),
\label{vector}
\end{align}
where $r=r(y)$ denotes the radius of the 
lattice plane whose $y$-coordinate is $y$ 
in the original system, and 
$u=u(x,y)$ and $v=v(x,y)$ are small deviations 
around the complete ring shape 
({\it i.e.}, dislocation-free ring). 
We define the metric tensor $g_{ij}$ by 
$(d \vec{R})^2=\sum_{i,j}g_{ij} dx_i dx_j$, 
where 
$g_{ij}=\frac{\partial \vec{R}}{\partial x_i} 
\cdot \frac{\partial \vec{R}}{\partial x_j}$
($i,j=\{1,2\}$, and $x_1=x$, $x_2=y$). 
The elements of the metric tensor are given as 
\begin{align}
g_{11}\equiv g_{xx}&=\left[\frac{r+v}{R}(1+u_x)^2\right]^2
+\left[v_x\right]^2,\\
g_{22}\equiv g_{yy}&=\left[r'+v_y \right]^2
+\left[\frac{u_y}{R}(r+v)\right]^2,\\
g_{12}\equiv g_{xy}&=\frac{1}{R^2} (r+v)^2 (1+u_x) u_y
+v_x (r'+v_y),
\end{align}
where $r'=d r/d y$, $u_x=\partial u/\partial x \cdots$ etc.
In this paper, we assume 
$r(y) = R + y$ ($-W/2 \le y \le W/2$) 
for simplicity. 
Strain tensor $u_{ij}$, 
which describes the elastic deformation of the crystal, 
is given by
$u_{ij}=\frac{1}{2}(g_{ij}-\delta_{ij})$, 
where $\delta_{ij}$ is the Kronecker's delta. 
Assuming an orthorhombic system for simplicity, 
the elastic free energy is given by 
\begin{align}
F=\int_{-L/2}^{L/2} & dx \int_{-W/2}^{W/2} dy\, 
\biggl\{\frac{1}{2}\lambda_{xxxx} (u_{xx}^2)
+\frac{1}{2}\lambda_{yyyy} (u_{yy}^2)\nonumber\\
&+\lambda_{xxyy} u_{xx} u_{yy}
+2 \lambda_{xyxy} (u_{xy})^2
\biggr\},
\label{free_en}
\end{align}
where $L=2 \pi R$ is the circumference of the ring
\cite{Landau}. 
We also assume that the principal axes of the crystal 
are along $x$- and $y$-axis, 
which yields $\lambda_{xxyy}=-\lambda_{xyxy}/2$. 
From now on we 
denote $\lambda_1 \equiv \lambda_{xxxx}$, 
$\lambda_2 \equiv \lambda_{yyyy}$ 
and $\lambda_3\equiv\lambda_{xyxy} =- \lambda_{xxyy}/2$.

Next we determine the free energy of the ring crystal within the linear 
elasticity. 
To do this we expand the free energy with respect to 
small parameters, 
$u_x$, $u_y$, $v_x$, $v_y$, $y/R$ and $v/R$ 
to the second order, which yields 
\begin{align}
F=\int d\vec{r}
\biggl[ &\frac{\lambda_1 }{2}
\left(\frac{y}{R}+ u_x+\frac{1}{R}v\right)^2
\nonumber+\frac{\lambda_2 }{2} v_y^2 +\frac{\lambda_3}{2} (u_y+v_x)^2 \phantom{\biggl[}
\nonumber\\
&- \frac{\lambda_3}{2}
\left(\frac{y}{R}+ u_x+\frac{1}{R}v\right) v_y
\biggr].\label{lin_ela}
\end{align}
Here we limit ourselves to the 
crystals with $W \ll R$. 
We further neglect $v$-field here, 
since $\lambda_2 \gg \lambda_3$ in most ring crystals, 
and dominant elastic fluctuation is expected to arise 
not from $v$ but from $u$. 
Then we obtain the following free energy, 
\begin{align}
F=&\int d\vec{r}
\biggl[ \frac{\lambda_1 }{2}
\left(u_x+\frac{y}{R}\right)^2
+\frac{\lambda_3}{2} u_y^2 
\biggr]. \label{lin_ela2}
\end{align}
By redefining the parameters by 
\begin{align}
& \gamma=\sqrt{\frac{\lambda_3}{\lambda_1}},\,\,\,\,
K=\frac{\lambda_1\gamma d_x^2}{(2 \pi)^2}, \,\,\,\,\,
\bar{y}=\frac{y}{\gamma},\,\,\,
\overline{W}=\frac{W}{\gamma},\nonumber\\
&A=-\frac{\phi_0 \gamma}{d_x R}\bar{y},\,\,\,
\theta(x,\bar{y})=\frac{2 \pi u(x,y)}{d_x}, 
\label{scale_param}
\end{align}
$F$ is rewritten as  
\begin{align}
F=\int_{-L/2}^{L/2} dx \int_{-\overline{W}/2}^{\overline{W}/2} d \bar{y}\,
\frac{K}{2} 
\left[\left(\theta_x-\frac{2\pi}{\phi_0}A\right)^2
+\theta_{\bar{y}}^2\right], 
\label{free_en_phase}
\end{align}
where $d_x$ and $\phi_0$ are the lattice constant in $x$-direction 
and the superconducting magnetic flux quantum, respectively. 
This is nothing but the free energy of an 
isotropic superconductor\cite{Tinkham}. 
The superconducting \lq\lq phase\rq\rq\, $\theta$ undergoes 
the quantization condition 
$\int_C d\vec{l}\cdot\nabla\theta = 2 \pi n_v$, 
where $C$ is a simple loop and 
$n_v$ is the total number of vortices enclosed 
in $C$. 
The vortices correspond to the edge dislocations in the 
crystal system. 
Thus the quantization of vorticity corresponds to 
the quantization of the Burgers vector. 
Here the vector potential $A$ is proportional to $1/R$, 
which means that the strength of 
\lq\lq geometrical\rq\rq\, frustration 
is proportional to the curvature. 

\vspace{2mm}
\noindent
\underline{\it Entry of the first dislocation}

\vspace{1mm}
Here we estimate the free energy of a single dislocation 
and study the condition for the creation of the first 
dislocation in the system as the strength of frustration 
is increased. 
We should note that 
the \lq\lq penetration depth\rq\rq\, 
$\lambda$ is infinite in the present system. 
In a $\lambda=\infty$ (extremely type-II) superconductor, 
as in the case of rotating superfluid, 
the lower critical field $H_{\rm c1}$ depends on 
the system size \cite{Vinen}. 
We encounter a similar situation here. 

We assume that the dislocation is 
located at the center of the system ($x=0$, $\bar{y}=0$) 
and, instead of treating the boundary condition 
seriously, we employ a rather simple cutoff procedure 
\cite{Hayashi-Yoshioka}. 
The phase modulation due to the vortex 
is put as 
$\nabla \theta=-\hat{z}\times \vec{r}/|\vec{r}|^2$, 
where $\hat{z}$ is the unit vector in $z$-direction 
and $\vec{r} = (x,{\bar y})$. 
Substituting this into Eq. (\ref{free_en_phase}), 
we obtain the dislocation free energy 
$F_{\rm d}  =  K \pi \log \frac{\overline{W}}{\bar{d}_y}
-\frac{\pi^2 K\gamma}{2 d_x R} \overline{W}^2$. 
Here the divergence at $\bar{y}=0$ is cutoff at 
$|\bar{y}|=\bar{d}_y=d_y/\gamma$ ($d_y$ is 
the lattice constant in $y$-direction). 
The condition for the dislocation creation
(namely, $F_{\rm d} < 0$) is, then, obtained as
\begin{align}
R< R_{c1} = \frac{\pi}{2 \gamma d_x}\frac{W^2}{\ln(W/d_y)}. 
\label{cond1}
\end{align}

\vspace{2mm}
\noindent
\underline{\it Many dislocations and destruction of elasticity}

\vspace{1mm}
Now we study the strongly frustrated regime where 
many dislocations are created. 
In this case, the size of the dislocation \lq\lq core\rq\rq\, 
is important. 
To study this, we discretize again the $\bar y$-coordinate 
in Eq. (\ref{lin_ela2}) as follows, 
\begin{align}
F &= d_y \int_0^L dx\,
\biggl[\sum_{j=1}^{N_y} 
\frac{\lambda_1}{2}\left(\partial_x u_j + \frac{y}{R}\right)^2
\nonumber\\
&-\sum_{j=1}^{N_y-1}
\frac{\lambda_3}{4 \pi^2}
\left(\frac{d_x}{d_y}\right)^2
\cos \left\{\frac{2 \pi (u_{j+1}-u_j)}{d_x}\right\}\biggr]. 
\label{fe}
\end{align}
Because of large anisotropy $\lambda_1 \gg \lambda_3$, 
the core is strongly elongated along $x$-direction 
although the size in $y$-direction is comparable to the 
lattice spacing. 
Therefore we extract two crystal planes 
between which the core is located and 
neglect the coupling to other planes. 
The solution which minimizes $F$ in Eq. (\ref{fe}) can be obtained 
analytically as 
$u_{j+1}=-u_j=(d_x/\pi) \arctan (x/\Lambda)$ 
(the core is locate between the $j$-th and the $(j+1)$-th plane). 
The core size in $x$-direction, $\Lambda$, is given by 
$\Lambda=d_y/\gamma$ (see Ref. \cite{Hayashi_cdw_dislo}
for a similar calculation). 
If we consider that the core size in $y$-direction to be 
$d_y$, the core size in the scaled system 
(see Eq. (\ref{scale_param})) is given by $d_y/\gamma$ 
in both $x$- and $y$-direction 
and the core is considered to be isotropic. 

In case of a type-II superconductor, 
the order is destroyed by a magnetic field when the 
distance between vortices becomes comparable to the 
vortex core radius, and this field strength is 
called the upper critical field 
$H_{c2} = \phi_0 /(2 \pi \xi^2)$, where $\xi$ is the coherence length. 
In the present case, 
noting the correspondences, 
\begin{align}
\frac{H}{\phi_0} \Leftrightarrow \frac{\gamma}{d_x R}, \,\,\,\, 
\xi\Longleftrightarrow \frac{d_y}{\gamma},
\end{align}
where $H$ is the magnetic field, 
we obtain the critical radius $R_{c2} = 2\pi d_y^2/(\gamma d_x)$ 
corresponding to $H_{c2}$. 
If $R < R_{c2}$, the frustration due to bending is 
so strong that the crystal is full of dislocations, 
whose cores are almost overlapped. 
In this case, each plane can be almost free from the 
elastic coupling to neighboring planes. 
In other words, the inter-plane 
elasticity is destroyed by the dislocations. 
If $R_{c2} < R < R_{c1}$, 
the density of dislocations is low enough and the cores are 
separated from each other. 
In this case, the crystal may retain a finite elasticity. 

\vspace{2mm}
\noindent
\underline{\it Phase diagram of ring crystals}

\begin{figure}[htb]
\includegraphics[width=7cm,clip]{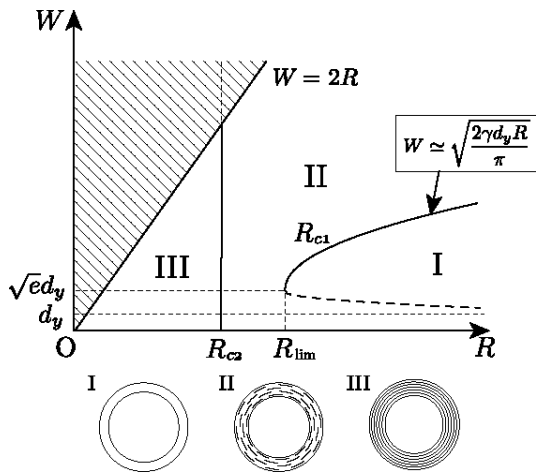}
    \caption{
The phase diagram of a ring crystal 
and the dislocation pattern in each phase (see text). 
In the crystal pictures, dislocation cores are shown by 
parts of arcs. }
    \label{pd}
\end{figure}

\vspace{1mm}
The $R$-$W$ phase diagram of ring crystals is 
depicted in Fig. \ref{pd}. 
As is simply imagined, $R > W/2$ is an 
absolute requirement for the formation of a ring crystal. 
The forbidden area is hatched in the figure. 
It should also be noted that 
our approach is not reliable in 
the region too close to $R = W/2$, 
because of the assumption $R \gg W$.

We have divided the whole region into three \lq\lq phases\rq\rq 
(I $\sim$ III). 
In the phase I, the crystal is free from dislocations 
(the \lq\lq Meissner phase\rq\rq\, in superconducting analog). 
The phase boundary $R_{c1}$ 
is reentrant as a function of $W$ 
because of the logarithmic contribution 
in Eq. (\ref{cond1}), 
although the lower part of the curve is meaningless because 
$W \le \sqrt{e} d_y$. 
(We denote the smallest $R$ in the phase I by $R_{\rm lim}$.) 
In the phase II, the frustration is not too large and 
the system contains dislocations well separated from each other 
(the \lq\lq mixed state\rq\rq), 
and the crystal may still hold rigidity. 
In the phase III, the frustration is large and the 
dislocation cores are overlapped, destroying the 
inter-plane rigidity of the crystal 
(the \lq\lq normal state\rq\rq). 
Possible dislocation pattern in each phase is also depicted in 
Fig. \ref{pd}.

The distribution of dislocations are observed directly 
by scanning tunneling microscope etc. 
Phonon spectrum measurement is also a 
powerful tool to investigate the crystal rigidity, 
since there are drastic changes in elasticity among 
I $\sim$ III. 
This may affect the physical properties of the system also. 
Studies on various ring crystals 
from such point of view may be an interesting topic 
in the future \cite{Tanda_pc,Nogawa}. 
Actually, Tanda et al. have 
observed that the critical temperature of charge density wave 
transition in NbSe$_3$ is significantly modified in ring 
crystals, whose relation to crystal defects may be worth investigating.

Now we apply our theory to actual systems. 
First, we study multi-wall carbon nanotubes, 
whose elastic constants are approximately those of graphite. 
From Ref. \cite{Blakslee}, we read 
$\lambda_1=c_{11}\simeq 100$, 
$\lambda_2=c_{33}\simeq 4$, 
$\lambda_3=c_{44}\simeq 0.02$ (dyn/cm$^2$), 
and $\lambda_3/\lambda_1 = \gamma^2 \simeq 10^{-4}$. 
Using $d_x = 0.25$ nm and $d_y = 0.34$ nm,  
we obtain $R_{c2}=46.2$ nm and $R_{\rm lim} = 395$ nm. 
When $R < R_{c2}$, the crystal is 
full of dislocations, and the elastic coupling between 
graphite sheets is negligible. 
This is consistent with the fact that 
thin nanotubes consist of graphite sheets with different chiralities 
and the inter-plane coupling becomes significant only in thicker tubes 
($R > 75$nm)\cite{Maniwa}. 
However an accurate treatment of different chiralities 
based on dislocation theory is a future problem. 

In case of transition metal trichargogenides, 
we study orthorhombic TaS$_3$, for which 
the order of $\lambda_1/\lambda_2$ is estimated 
from Young modulus $E$ and shear modulus $G$ 
in Ref. \cite{Xiang-Brill,Yamaya-Oomi} 
as $G/E =\lambda_3/\lambda_1 = \gamma^2
\simeq 10^{-2}$. 
Using $d_x = 0.3$ nm and 
$d_y = 2$ nm, 
we obtain,  $R_{c2}=133.3$ nm and 
$R_{\rm lim} = 1138$ nm. 
Although the phase I may be difficult to realize 
(e.g., $R\simeq 30\mu$m and 
$W\simeq 40$nm), crossover from II to III 
may be experimentally accessible. 

It is also interesting to investigate tubes with $R \simeq W$ 
by numerically minimizing the free energy, 
Eq. (\ref{free_en}) or (\ref{lin_ela}), 
instead of using superconducting analog. 
Such study is now prepared. 

In summary, we have clarified the distribution of dislocations, 
in the ring-(or cylinder-)shaped crystals with various 
radiuses and thicknesses. 
Our results may be useful for the studies of the 
physical properties of \lq\lq topological matters\rq\rq, 
and for their future applications to electrical and 
mechanical devices. 

MH and HE were financially supported by a Grant-In-Aid for 
Scientific Research from the Ministry of 
Education, Science, Sports and Culture, Japan. 
KK was financially supported by the Sumitomo Foundation.

\vfill 

\begin{thebibliography}{29}
\expandafter\ifx\csname natexlab\endcsname\relax\def\natexlab#1{#1}\fi
\expandafter\ifx\csname bibnamefont\endcsname\relax
  \def\bibnamefont#1{#1}\fi
\expandafter\ifx\csname bibfnamefont\endcsname\relax
  \def\bibfnamefont#1{#1}\fi
\expandafter\ifx\csname citenamefont\endcsname\relax
  \def\citenamefont#1{#1}\fi
\expandafter\ifx\csname url\endcsname\relax
  \def\url#1{\texttt{#1}}\fi
\expandafter\ifx\csname urlprefix\endcsname\relax\def\urlprefix{URL }\fi
\providecommand{\bibinfo}[2]{#2}
\providecommand{\eprint}[2][]{\url{#2}}

\bibitem[{\citenamefont{Ashcroft and Mermin}(1976)}]{Ashcroft-Mermin}
\bibinfo{author}{\bibfnamefont{N.~W.} \bibnamefont{Ashcroft}} \bibnamefont{and}
  \bibinfo{author}{\bibfnamefont{N.~D.} \bibnamefont{Mermin}},
  \emph{\bibinfo{title}{Solid State Physics}} (\bibinfo{publisher}{Saunders
  college}, \bibinfo{year}{1976}).

\bibitem[{\citenamefont{Iijima}(1987)}]{iijima}
\bibinfo{author}{\bibfnamefont{S.}~\bibnamefont{Iijima}}, \bibinfo{journal}{J.
  Phys. Chem.} p. \bibinfo{pages}{3466} (\bibinfo{year}{1987}).

\bibitem[{\citenamefont{Curl and Smally}(1988)}]{curl-smally}
\bibinfo{author}{\bibfnamefont{R.~F.} \bibnamefont{Curl}} \bibnamefont{and}
  \bibinfo{author}{\bibfnamefont{R.~E.} \bibnamefont{Smally}},
  \bibinfo{journal}{Science} \textbf{\bibinfo{volume}{242}},
  \bibinfo{pages}{1017} (\bibinfo{year}{1988}).

\bibitem[{\citenamefont{Kroto}(1988)}]{kroto_review}
\bibinfo{author}{\bibfnamefont{H.~W.} \bibnamefont{Kroto}},
  \bibinfo{journal}{Science} \textbf{\bibinfo{volume}{242}},
  \bibinfo{pages}{1139} (\bibinfo{year}{1988}).

\bibitem[{\citenamefont{Iijima}(1991)}]{iijima_nanotube}
\bibinfo{author}{\bibfnamefont{S.}~\bibnamefont{Iijima}},
  \bibinfo{journal}{Nature} \textbf{\bibinfo{volume}{354}}, \bibinfo{pages}{56}
  (\bibinfo{year}{1991}).

\bibitem[{\citenamefont{Saito et~al.}(1998)\citenamefont{Saito, Dresselhaus,
  and Dresselhaus}}]{saito}
\bibinfo{author}{\bibfnamefont{R.}~\bibnamefont{Saito}},
  \bibinfo{author}{\bibfnamefont{G.}~\bibnamefont{Dresselhaus}},
  \bibnamefont{and} \bibinfo{author}{\bibfnamefont{M.~S.}
  \bibnamefont{Dresselhaus}}, \emph{\bibinfo{title}{Physical Properties of
  Carbon Nanotubes}} (\bibinfo{publisher}{Imperial College Press},
  \bibinfo{year}{1998}).

\bibitem[{\citenamefont{Hayashi}(2005)}]{hayashi-nanotube}
\bibinfo{author}{\bibfnamefont{M.}~\bibnamefont{Hayashi}},
  \bibinfo{journal}{Phy. Lett. A} \textbf{\bibinfo{volume}{342}},
  \bibinfo{pages}{237} (\bibinfo{year}{2005}).

\bibitem[{\citenamefont{Tanda et~al.}(1999)\citenamefont{Tanda, Kawamoto,
  Shiobara, Okajima, and Yamaya}}]{Tanda1}
\bibinfo{author}{\bibfnamefont{S.}~\bibnamefont{Tanda}},
  \bibinfo{author}{\bibfnamefont{H.}~\bibnamefont{Kawamoto}},
  \bibinfo{author}{\bibfnamefont{M.}~\bibnamefont{Shiobara}},
  \bibinfo{author}{\bibfnamefont{Y.}~\bibnamefont{Okajima}}, \bibnamefont{and}
  \bibinfo{author}{\bibfnamefont{K.}~\bibnamefont{Yamaya}},
  \bibinfo{journal}{J. Phys. IV France} \textbf{\bibinfo{volume}{9}},
  \bibinfo{pages}{379} (\bibinfo{year}{1999}).

\bibitem[{\citenamefont{Tanda et~al.}(2000)\citenamefont{Tanda, Kawamoto,
  Shiobara, Sakai, Yasuzuka, Okajima, and Yamaya}}]{tanda2}
\bibinfo{author}{\bibfnamefont{S.}~\bibnamefont{Tanda}},
  \bibinfo{author}{\bibfnamefont{H.}~\bibnamefont{Kawamoto}},
  \bibinfo{author}{\bibfnamefont{M.}~\bibnamefont{Shiobara}},
  \bibinfo{author}{\bibfnamefont{Y.}~\bibnamefont{Sakai}},
  \bibinfo{author}{\bibfnamefont{S.}~\bibnamefont{Yasuzuka}},
  \bibinfo{author}{\bibfnamefont{Y.}~\bibnamefont{Okajima}}, \bibnamefont{and}
  \bibinfo{author}{\bibfnamefont{K.}~\bibnamefont{Yamaya}},
  \bibinfo{journal}{Physica B} \textbf{\bibinfo{volume}{284-288}},
  \bibinfo{pages}{1657} (\bibinfo{year}{2000}).

\bibitem[{\citenamefont{Okajima et~al.}(2000)\citenamefont{Okajima, Kawamoto,
  Shiobara, Matsuda, Tanda, and Yamaya}}]{Okajima}
\bibinfo{author}{\bibfnamefont{Y.}~\bibnamefont{Okajima}},
  \bibinfo{author}{\bibfnamefont{H.}~\bibnamefont{Kawamoto}},
  \bibinfo{author}{\bibfnamefont{M.}~\bibnamefont{Shiobara}},
  \bibinfo{author}{\bibfnamefont{K.}~\bibnamefont{Matsuda}},
  \bibinfo{author}{\bibfnamefont{S.}~\bibnamefont{Tanda}}, \bibnamefont{and}
  \bibinfo{author}{\bibfnamefont{K.}~\bibnamefont{Yamaya}},
  \bibinfo{journal}{Physica B} \textbf{\bibinfo{volume}{284-288}},
  \bibinfo{pages}{1659} (\bibinfo{year}{2000}).

\bibitem[{\citenamefont{Tanda et~al.}(2002)\citenamefont{Tanda, Tsuneta,
  Okajima, Inagaki, Yamaya, and Hatakenaka}}]{Tanda}
\bibinfo{author}{\bibfnamefont{S.}~\bibnamefont{Tanda}},
  \bibinfo{author}{\bibfnamefont{T.}~\bibnamefont{Tsuneta}},
  \bibinfo{author}{\bibfnamefont{Y.}~\bibnamefont{Okajima}},
  \bibinfo{author}{\bibfnamefont{K.}~\bibnamefont{Inagaki}},
  \bibinfo{author}{\bibfnamefont{K.}~\bibnamefont{Yamaya}}, \bibnamefont{and}
  \bibinfo{author}{\bibfnamefont{N.}~\bibnamefont{Hatakenaka}},
  \bibinfo{journal}{Nature} \textbf{\bibinfo{volume}{417}},
  \bibinfo{pages}{397} (\bibinfo{year}{2002}).

\bibitem[{\citenamefont{Tsuneta and Tanda}(2004)}]{Tsuneta}
\bibinfo{author}{\bibfnamefont{T.}~\bibnamefont{Tsuneta}} \bibnamefont{and}
  \bibinfo{author}{\bibfnamefont{S.}~\bibnamefont{Tanda}}, \bibinfo{journal}{J.
  Crystal Growth} \textbf{\bibinfo{volume}{264}}, \bibinfo{pages}{223}
  (\bibinfo{year}{2004}).

\bibitem[{\citenamefont{Seifert et~al.}(2000)\citenamefont{Seifert, Terrones,
  Terrones, Jungnickel, and Frauenheim}}]{MoS2}
\bibinfo{author}{\bibfnamefont{G.}~\bibnamefont{Seifert}},
  \bibinfo{author}{\bibfnamefont{H.}~\bibnamefont{Terrones}},
  \bibinfo{author}{\bibfnamefont{M.}~\bibnamefont{Terrones}},
  \bibinfo{author}{\bibfnamefont{G.}~\bibnamefont{Jungnickel}},
  \bibnamefont{and}
  \bibinfo{author}{\bibfnamefont{T.}~\bibnamefont{Frauenheim}},
  \bibinfo{journal}{Phys. Rev. Lett.} \textbf{\bibinfo{volume}{85}},
  \bibinfo{pages}{146} (\bibinfo{year}{2000}).

\bibitem[{\citenamefont{Gr{\"u}ner}(1994)}]{Gruner}
\bibinfo{author}{\bibfnamefont{G.}~\bibnamefont{Gr{\"u}ner}},
  \emph{\bibinfo{title}{Density Waves in Solids}}
  (\bibinfo{publisher}{Addison-Wesley}, \bibinfo{year}{1994}).

\bibitem[{\citenamefont{Hayashi and Ebisawa}(2002)}]{Hayashi-Ebisawa}
\bibinfo{author}{\bibfnamefont{M.}~\bibnamefont{Hayashi}} \bibnamefont{and}
  \bibinfo{author}{\bibfnamefont{H.}~\bibnamefont{Ebisawa}},
  \bibinfo{journal}{J. Phys. Soc. Jpn.} \textbf{\bibinfo{volume}{70}},
  \bibinfo{pages}{3495} (\bibinfo{year}{2002}).

\bibitem[{\citenamefont{Hayashi et~al.}(2005)\citenamefont{Hayashi, Ebisawa,
  and Kuboki}}]{Hayashi-Ebisawa-Kuboki}
\bibinfo{author}{\bibfnamefont{M.}~\bibnamefont{Hayashi}},
  \bibinfo{author}{\bibfnamefont{H.}~\bibnamefont{Ebisawa}}, \bibnamefont{and}
  \bibinfo{author}{\bibfnamefont{K.}~\bibnamefont{Kuboki}},
  \bibinfo{journal}{Phys. Rev. B} \textbf{\bibinfo{volume}{72}},
  \bibinfo{pages}{024505} (\bibinfo{year}{2005}).

\bibitem[{\citenamefont{Yakubo et~al.}(2003)\citenamefont{Yakubo, Avishai, and
  Cohen}}]{Yakubo-Avishai-Cohen}
\bibinfo{author}{\bibfnamefont{K.}~\bibnamefont{Yakubo}},
  \bibinfo{author}{\bibfnamefont{Y.}~\bibnamefont{Avishai}}, \bibnamefont{and}
  \bibinfo{author}{\bibfnamefont{D.}~\bibnamefont{Cohen}},
  \bibinfo{journal}{Phys. Rev. B} \textbf{\bibinfo{volume}{67}},
  \bibinfo{pages}{125319} (\bibinfo{year}{2003}).

\bibitem[{\citenamefont{Wakabayashi and Harigaya}(2003)}]{Wakabayashi-Harigaya}
\bibinfo{author}{\bibfnamefont{K.}~\bibnamefont{Wakabayashi}} \bibnamefont{and}
  \bibinfo{author}{\bibfnamefont{K.}~\bibnamefont{Harigaya}},
  \bibinfo{journal}{J. Phys. Soc. Jpn.} \textbf{\bibinfo{volume}{72}},
  \bibinfo{pages}{998} (\bibinfo{year}{2003}).

\bibitem[{\citenamefont{Landau and Lifshitz}(1986)}]{Landau}
\bibinfo{author}{\bibfnamefont{L.~D.} \bibnamefont{Landau}} \bibnamefont{and}
  \bibinfo{author}{\bibfnamefont{E.~M.} \bibnamefont{Lifshitz}},
  \emph{\bibinfo{title}{Theory of Elasticity}} (\bibinfo{publisher}{Pergamon,
  New York}, \bibinfo{year}{1986}).

\bibitem[{\citenamefont{Tinkham}(1996)}]{Tinkham}
\bibinfo{author}{\bibfnamefont{M.}~\bibnamefont{Tinkham}},
  \emph{\bibinfo{title}{Superconductivity}} (\bibinfo{publisher}{McGraw-Hill},
  \bibinfo{year}{1996}).

\bibitem[{\citenamefont{Vinen}()}]{Vinen}
\bibinfo{author}{\bibfnamefont{W.~F.} \bibnamefont{Vinen}},
  \bibinfo{note}{chapt. 20 of {\it Superconductivity}, R. D. Parks ed., (Marcel
  Dekker, 1969).}

\bibitem[{\citenamefont{Hayashi and Yoshioka}(1996)}]{Hayashi-Yoshioka}
\bibinfo{author}{\bibfnamefont{M.}~\bibnamefont{Hayashi}} \bibnamefont{and}
  \bibinfo{author}{\bibfnamefont{H.}~\bibnamefont{Yoshioka}},
  \bibinfo{journal}{Phys. Rev. Lett.} \textbf{\bibinfo{volume}{77}},
  \bibinfo{pages}{3403} (\bibinfo{year}{1996}).

\bibitem[{\citenamefont{Hayashi}(2002)}]{Hayashi_cdw_dislo}
\bibinfo{author}{\bibfnamefont{M.}~\bibnamefont{Hayashi}},
  \bibinfo{journal}{Physica B} \textbf{\bibinfo{volume}{324}},
  \bibinfo{pages}{82} (\bibinfo{year}{2002}).

\bibitem[{\citenamefont{Shimatake et~al.}()\citenamefont{Shimatake, Toda, and
  Tanda}}]{Tanda_pc}
\bibinfo{author}{\bibfnamefont{K.}~\bibnamefont{Shimatake}},
  \bibinfo{author}{\bibfnamefont{Y.}~\bibnamefont{Toda}}, \bibnamefont{and}
  \bibinfo{author}{\bibfnamefont{S.}~\bibnamefont{Tanda}},
  \bibinfo{note}{cond-mat/0511328}.

\bibitem[{\citenamefont{Nogawa and Nemoto}()}]{Nogawa}
\bibinfo{author}{\bibfnamefont{T.}~\bibnamefont{Nogawa}} \bibnamefont{and}
  \bibinfo{author}{\bibfnamefont{K.}~\bibnamefont{Nemoto}},
  \bibinfo{note}{cond-mat/0511326}.

\bibitem[{\citenamefont{Blakslee et~al.}(1970)\citenamefont{Blakslee, Proctor,
  Seldin, Spence, and Weng}}]{Blakslee}
\bibinfo{author}{\bibfnamefont{O.~L.} \bibnamefont{Blakslee}},
  \bibinfo{author}{\bibfnamefont{D.~G.} \bibnamefont{Proctor}},
  \bibinfo{author}{\bibfnamefont{E.~J.} \bibnamefont{Seldin}},
  \bibinfo{author}{\bibfnamefont{G.~B.} \bibnamefont{Spence}},
  \bibnamefont{and} \bibinfo{author}{\bibfnamefont{T.}~\bibnamefont{Weng}},
  \bibinfo{journal}{J. Appl. Phys.} \textbf{\bibinfo{volume}{41}},
  \bibinfo{pages}{3373} (\bibinfo{year}{1970}).

\bibitem[{\citenamefont{Maniwa et~al.}(2001)\citenamefont{Maniwa, Fujiwara,
  Kira, Tou, Nishibori, Takata, Sakata, Fujiwara, Zhao, Iijima
  et~al.}}]{Maniwa}
\bibinfo{author}{\bibfnamefont{Y.}~\bibnamefont{Maniwa}},
  \bibinfo{author}{\bibfnamefont{R.}~\bibnamefont{Fujiwara}},
  \bibinfo{author}{\bibfnamefont{H.}~\bibnamefont{Kira}},
  \bibinfo{author}{\bibfnamefont{H.}~\bibnamefont{Tou}},
  \bibinfo{author}{\bibfnamefont{E.}~\bibnamefont{Nishibori}},
  \bibinfo{author}{\bibfnamefont{M.}~\bibnamefont{Takata}},
  \bibinfo{author}{\bibfnamefont{M.}~\bibnamefont{Sakata}},
  \bibinfo{author}{\bibfnamefont{A.}~\bibnamefont{Fujiwara}},
  \bibinfo{author}{\bibfnamefont{X.}~\bibnamefont{Zhao}},
  \bibinfo{author}{\bibfnamefont{S.}~\bibnamefont{Iijima}},
  \bibnamefont{et~al.}, \bibinfo{journal}{Phys. Rev. B}
  \textbf{\bibinfo{volume}{64}}, \bibinfo{pages}{073105}
  (\bibinfo{year}{2001}).

\bibitem[{\citenamefont{Xiang and Brill}(1987)}]{Xiang-Brill}
\bibinfo{author}{\bibfnamefont{X.~D.} \bibnamefont{Xiang}} \bibnamefont{and}
  \bibinfo{author}{\bibfnamefont{J.~W.} \bibnamefont{Brill}},
  \bibinfo{journal}{Phy. Rev. B} \textbf{\bibinfo{volume}{36}},
  \bibinfo{pages}{2969} (\bibinfo{year}{1987}).

\bibitem[{\citenamefont{Yamaya and Oomi}(1982)}]{Yamaya-Oomi}
\bibinfo{author}{\bibfnamefont{K.}~\bibnamefont{Yamaya}} \bibnamefont{and}
  \bibinfo{author}{\bibfnamefont{G.}~\bibnamefont{Oomi}}, \bibinfo{journal}{J.
  Phys. Soc. Jpn.} \textbf{\bibinfo{volume}{51}}, \bibinfo{pages}{3512}
  (\bibinfo{year}{1982}).

\end{thebibliography}
\end{document}